\documentclass[useAMS,usenatbib]{mn2e}
\usepackage[dvipdfmx]{graphicx}
\usepackage{amsmath,amssymb,natbib,siunitx,booktabs,url}
\usepackage{color}

\newcommand{\lsim}{\lesssim}
\newcommand{\gsim}{\gtrsim}
\newcommand{\psim}{\mbox{\raisebox{-1.0ex}{$~\stackrel{\textstyle \propto}
{\textstyle \sim}~$ }}}

\newcommand{\lmk}{\left(}
\newcommand{\rmk}{\right)}

\newcommand{\lla}{\left\langle}

\newcommand{\rra}{\right\rangle}
\newcommand{\so}{M_\odot}
\newcommand{\mch}{{\cal M}}

\begin{document}

\title[]{ How many extra-Galactic stellar-mass binary black holes will be detected by space gravitational-wave interferometers? }

\author[]
{Naoki Seto$^1$ and Koutarou Kyutoku$^{1,2,3}$\\
$^1$Department of Physics, Kyoto University, Kyoto 606-8502, Japan\\
$^2$Center for Gravitational Physics, Yukawa Institute for Theoretical Physics, Kyoto University, Kyoto 606-8502, Japan\\
$^3$Interdisciplinary Theoretical and Mathematical Sciences Program (iTHEMS), RIKEN, Wako, Saitama 351-0198, Japan
}

\date{\today}

\maketitle

\begin{abstract}

On the basis of  GWTC-3,  we discuss the detection prospect of   extra-Galactic binary black holes (BBHs) by space gravitational-wave interferometers.  In particular,  targeting BBHs with component masses around 5-100$\so$, we directly incorporate the chirp mass distribution of the 62 BBHs detected at high significance.   We find that, due to the reduction of both the comoving merger rate and a weighted average of  chirp masses, the expected detection numbers are generally much smaller than the results obtained by the same authors immediately after the report of GW150914.
For LISA, the total BBH detections are estimated to be $N_{\rm tot}\sim 2 (T/4{\rm yr})^{3/2}(\rho_{\rm thr}/10)^{-3}$, dominated by nearly monochromatic BBHs  ($\rho_{\rm thr}$: the detection threshold,  $T$: the observational period). 
TianQin will have a total detection number $N_{\rm tot}$ similar to LISA.   Meanwhile,  TianQin has potential to find $N_{\rm mer}\sim0.6  (T/4{\rm yr})^{7/4}(\rho_{\rm thr}/10)^{-3}$  BBHs that merge in the observational period. This number for merging BBHs is 4-5 times larger than that of  LISA, because of the difference between the optimal bands. We also investigate prospects for joint operations of multiple detectors, finding that concurrent observations will be more advantageous than sequential ones.
 
\end{abstract}


\section{introduction}
Since the discovery of the first event GW150914, Advanced LIGO and Advanced Virgo have observed mergers of $\sim 80$ binary black holes (BBHs), as recently summarized in the catalog  GWTC-3 (Abbott et al. 2021a).  The large number of  samples allows us to examine various astronomical properties such as the mass distribution of BBHs, the comoving merger rate  and its redshift dependence (Abbott et al. 2021b). 

BBHs are interesting observational  targets also for space interferometers  such as LISA (Amaro-Seoane et al. 2017).  The detection prospects by these future detectors can be evaluated by combining  astrophysical as well as instrumental information (noise curve, operation period etc.). In 2016, shortly after the announcement of GW150914 (Abbott et al. 2016), the authors studied the expected number of BBH detections with {LISA} (Kyutoku \& Seto 2016, paper 1, see also Sesana 2016; Del Pozzo, Sesana, \& Klein 2018; Liu et al. 2020).
In the past 6 years after the study, along with the significant improvement in our knowledge on BBHs,  the sensitivity goal of LISA was refined and narrowed down. Furthermore, two Chinese missions, TianQin and Taiji have been actively developed (Luo et al. 2016; Ruan et al. 2018). Therefore, comparing with paper 1, we can now draw a more precise picture about the prospects of the BBH observation with space interferometers.

 In this paper, by using GWTC-3, we estimate the total numbers of BBH detections with LISA and TianQin.  To this end,  we directly leverage  the mass distribution of the 62 BBHs detected so far at high significance level.  We also evaluate the numbers of BBHs that merge in operation periods of the detectors. The latter would be important for multi-band  observations (see e.g., Sesana 2016; Nair, Jhingan \& Tanaka 2016; Isoyama, Nakano \& Nakamura 2018).  We find that, due to reduction of  both the comoving merger rate and the weighted averaged of chirp masses, the expected numbers of BBH detections  become much smaller than the typical values presented in paper 1.

This paper is organized as follows. In \S 2, from GWTC-3, we extract information of BBHs relevant to this paper.  In \S 3,  we review the formulation in paper 1 for calculating the detectable comoving volume.  In \S 4, we evaluate the detectable numbers, using a simplified model and a mass-weighted model.  We also discuss the impacts of the joint operation of two space interferometers such as the LISA-TianQin pair. In \S 5, we make a brief discussion on possible modifications to our results. \S 6 is devoted to a summary of this paper.

\section{chirp mass distribution}

In Fig. 1, with the black dots, we show the observed (median) chirp masses $\mch_l$ (in the source frame) for the 62 BBHs reported in GWTC-3 with the false alarm rates (FARs) lower than 0.25yr$^{-1}$  (Abbott et al. 2021b).  For each event, we chronologically assign the label $l$ with $l=1$ for GW150914 and $l=62$ for GW200316\_21576. The maximum value $\mch_{20}=69.2\so$ is for GW190521\_030229.  We also show the beginnings of the O3a and O3b runs.  {Following Abbott et al. (2021b), we exclude the outlier GW190814 from our sample (closely related to the argument about the lower mass gap).  Therefore, we study BBHs with component masses in the range $[5\so, 100\so]$ and the chirp masses around $[5\so,70\so]$.   }

Abbott et al. (2021b) estimated 
the comoving merger rate  of BBHs in the range   $17.3-45 {\rm Gpc^{-3} yr^{-1}}$ at the redshift  $z=0.2$ with the fitted dependence $(1+z)^\kappa$ ($\kappa=2.7^{+1.8}_{-1.9}$).  Since LISA can only observe stellar mass BBHs at lower redshifts than current ground-based detectors, we estimate the mean rate  at $z=0$ by 
\begin{eqnarray}
R_0=(17.3+45)/2\times (1+0.2)^{-2.7}=19.1 {\rm Gpc^{-3} yr^{-1}},
\end{eqnarray}
expecting { an uncertainty factor of two (see Fig. 13 in Abbott et al. 2021b). }

\begin{figure}
 \includegraphics[width=.95\linewidth]{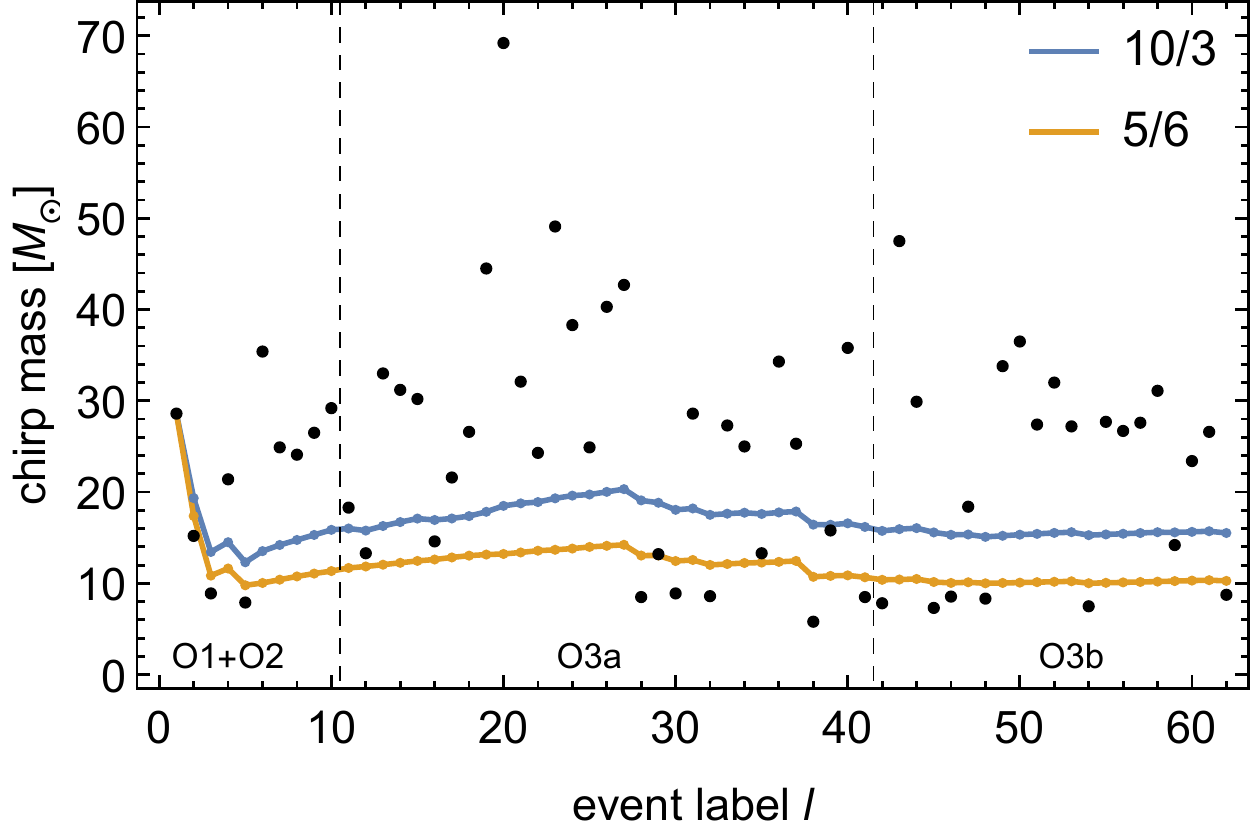} \caption{Observed (median) chirp masses $\mch_l$ (black dots) and the averaged ones ${\bar \mch}_\alpha$ with $\alpha=10/3$ and 5/6.  The total number of BBHs is 62  labelled by $l$ in the chronological order (O1: 3, O2: 7,  O3a: 31 and O3b : 21 events).  
 The weighted average of the chrip masses ${\bar \mch}_\alpha$ were evaluated by using the first $l$ events with   Eq. (\ref{mmnt}) .  
 }  \label{fig:volume}
\end{figure}

Next, we discuss the chirp mass distribution function $P(\mch)$ for the comoving merger rate, using the observed 62 chirp masses.   We impose the normalization 
\begin{equation}
\int_{\mch_{\rm min}}^{\mch_{\rm max}} d\mch P(\mch)=1
\end{equation}
in the mass range $[\mch_{\rm min}, \mch_{\rm max}]$.
In concrete terms, we simply assume that (i) a BBH can be  detected if and only if its signal-to-noise ratio exceeds a certain threshold, and (ii) the signal-to-noise ratio is dominated by the inspiral waveform ({$\propto \mch^{5/6}/D$} with the distance $D$).  Then, ignoring the cosmological effects, the detectable volume is $\propto D^3\propto \mch^{5/2}$. 

For the sake of conceptual convenience for dealing with the probability measure, we temporarily divide the chirp masses into finite segments with the labels $\sigma=1,\cdots,\sigma_{\rm max}$. We put $\Delta \mch_\sigma$ and $\mch_\sigma$ for the width and the central value for the segment $\sigma$  (with $\mch_{\sigma=1}\sim \mch_{\rm min}$ and $\mch_{\sigma=\sigma_{\rm max}}\sim \mch_{\rm max}$).  Our final results do not depend on the detail of the segmentation.

For a segment $\sigma$, the expected detection number is given by 
\begin{equation}
n_{\sigma}=K P(\mch_{\sigma}) \Delta \mch_{\sigma} \mch_{\sigma}^{5/2}
\end{equation}
with the factor $K$ independent of the segment. 
We then have  
\begin{equation}
P(\mch_{\sigma}) \Delta \mch_{\sigma}=K^{-1} n_{\sigma} \mch_{\sigma}^{-5/2}
\end{equation}
or
\begin{equation}
P(\mch_{\sigma}) \Delta \mch_{\sigma}=\frac{n_{\sigma} \mch_{\sigma}^{-5/2}}{\sum_{\sigma=1}^{\sigma_{\rm max}}n_{\sigma} \mch_{\sigma}^{-5/2}},
\end{equation}
taking into account the normalization.

Now we can approximately take the statistical average with respect to the probability distribution  $P(\mch)$.  We can formally write down the weighted value of a function $g(\mch)$ by
\begin{eqnarray}
\lla  g(\mch) \rra&=&
\int_{\mch_{\rm min}}^{\mch_{\rm max}}g(\mch)P(\mch)d\mch .
\end{eqnarray}
Using the segmentation, we have 
\begin{eqnarray}
\lla  g(\mch) \rra&\simeq& \sum_{\sigma=1}^{\sigma_{\rm max}}g(\mch_{\sigma})P(\mch_{\sigma}) \Delta \mch_{\sigma} \\
&\simeq& \frac{\sum_{\sigma} g(\mch_{\sigma})n_{\sigma}\mch_{\sigma}^{-5/2}}{\sum_{\sigma}n_{\sigma} \mch_{\sigma}^{-5/2}}\\
&\simeq& \frac{\sum_l   g(\mch_l)\mch_l^{-5/2}}{\sum_l\mch_l^{-5/2}} .\label{sl}
\end{eqnarray}
In Eq. (\ref{sl}), we replace the segment label $\sigma$ by the event label $l$.  This expression is independent of the segmentation. 

As an example,  we define the mean chirp mass with the power index $\alpha$ by
\begin{eqnarray}
{\bar \mch}_\alpha&\equiv &\lla  \mch^\alpha \rra^{1/\alpha}\\
&=& \lmk    \frac{\sum_l   \mch_l^{\alpha-5/2}}{\sum_l\mch_l^{-5/2}}\rmk^{1/\alpha}. \label{mmnt}
\end{eqnarray}
For our analysis below, the weighted average of the chirp  masses ${\bar \mch}_{10/3}$ is particularly important, as explained later with  Eq.  (\ref{eq:mono}).  In Fig. 1, together with  ${\bar \mch}_{5/6}$ (see \S4.2), we show  ${\bar \mch}_{10/3}$ obtained by using the first $l$ events in Eq. (\ref{sl}).  At the right end ($l=62$), we have {${\bar \mch}_{10/3}=15.5\so$ and ${\bar \mch}_{5/6}=10.3\so$ {($\sim 4\%$ smaller than the results at $l=41$ corresponding to  the end of O3a)}.  
In addition to the 62 BBHs with  $\rm FAR <0.25  yr^{-1}$,  Abbott et al. (2021b) provided seven  more BBHs  with $\rm 0.25 yr^{-1} <FAR <1  yr^{-1}$ (excluding GW190917).  Even using  69 BBHs,  the two  weighted averages of the  masses ${\bar \mch}_{10/3}$ and ${\bar \mch}_{5/6}$ change only by $\sim$1\%.  

%If we exclude GW190521, we have ${\bar \mch}_{10/3}=16.4\so$ and ${\bar \mch}_{5/2}=14.4\so$.

We can also evaluate the cumulative chirp mass distribution  by  
\begin{equation}
P(<\mch_{\rm thr})\equiv \int_{\mch_{\rm min}}^{\mch_{\rm thr}}P(\mch
)d\mch=\lla \theta (\mch_{\rm th}-\mch)  \rra .
\end{equation}
In Fig. 2, we present the numerical results obtained by using all the 62 events (from O1, O2, O3a and O3b) and the 41 events  excluding O3b.  We can  clearly see the deficiency in the range 9-13$\so$.
We comment that, in contrast to the original distribution $P(\mch)$,  the appearance of the cumulative distribution $P(<\mch_{\rm thr})$ is less affected by the smoothing operation associated with the discreteness of the samples.

 Abbott et al. (2021b) provided an estimate for the function $P(\mch)$, based on the flexible mixture (FM) model framework\footnote{\url{https://zenodo.org/record/5655785#.YdPOn9tUu0q}}.
 {In Fig. 2, we present its cumulative profile that shows a reasonable agreement with our results.} 
  Using  their distribution function,  we also evaluated the mean chirp masses, and obtained  ${\bar \mch}_{10/3}=16.4\so$ and ${\bar \mch}_{5/6}=10.8\so$. These numerical values are  larger only  by $\sim 5\%$  than our simple evaluations.

\begin{figure}
 \includegraphics[width=.95\linewidth]{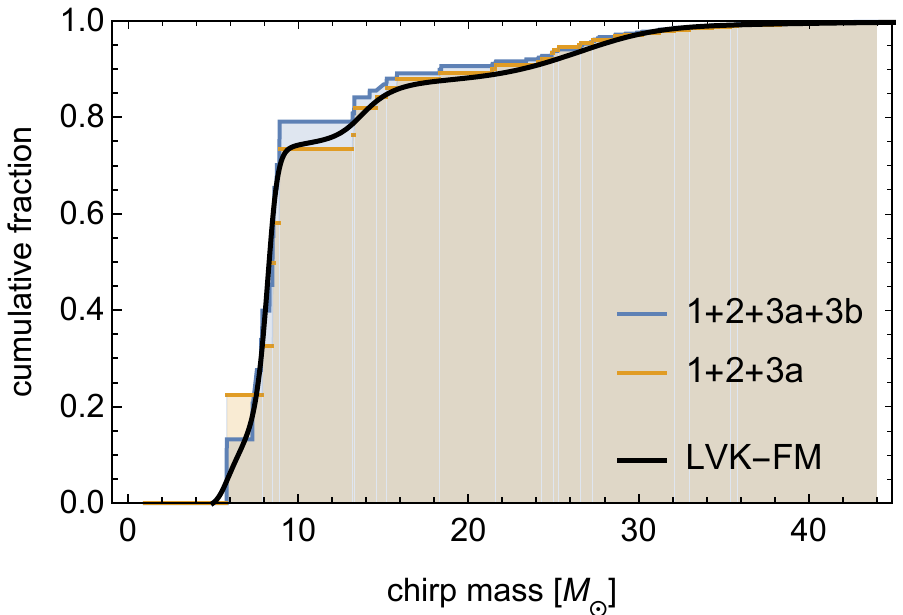} \caption{ The cumulative chrip mass distributions of BBHs for the comoving merger rate.  The orange steps are the  result from the 41 BBHs in O1+O2+O3a  and the blue steps are  from the 62 BBHs in O1+O2+O3a+O3b. The black curve is a model prediction in  Abbott et al. (2021b). } \label{fig:volume}
\end{figure}

In the analysis below, we use the comoving merger rate $R_0=19.1 {\rm Gpc^{-3} yr^{-1}}$ at $z=0$  and the direct  averaging (\ref{sl})  with the 62 BBH samples in the range $[\mch_{\rm min},\mch_{\rm max}]=[5\so,70\so]$,  resulting in   ${\bar \mch}_{10/3}=15.5\so$.   
In paper 1, we used the BBH  merger rate $R=100 {\rm Gpc^{-3} yr^{-1}}$ , assuming the single mass profile at $\mch=28.6\so$, mainly  based on  the first event GW150914  (Abbott et al. 2016).  These values are much larger than the updated ones in this paper.

\begin{figure}
 \includegraphics[width=.95\linewidth]{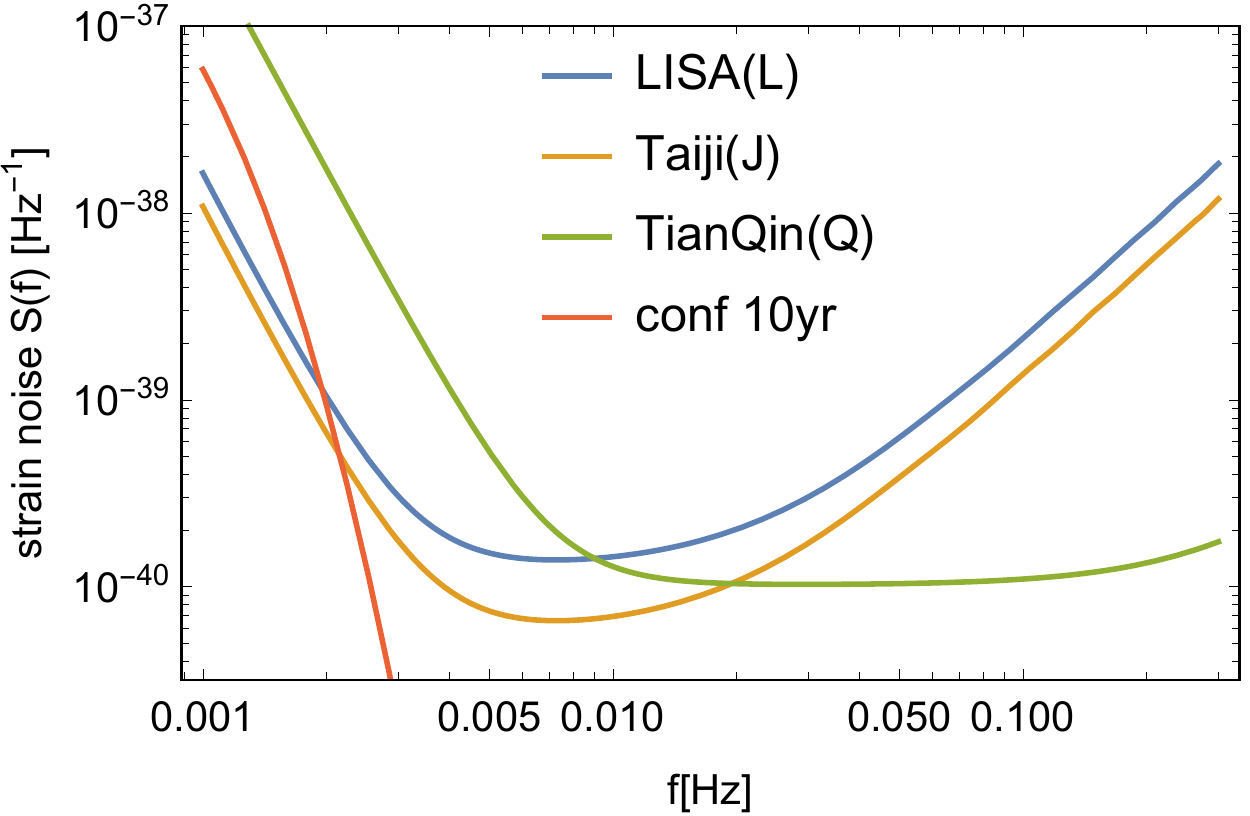} \caption{The angular averaged strain noise spectra $S(f)$ for LISA, Taiji and TianQin.  The Galactic confusion noise is given for $T=10$yr. }  \label{fig:volume}
\end{figure}

%\end{document}

\section{Evolution of Individual binary black holes}

In this section, following paper 1, we briefly summarize evolution of a circular BBH initially at a frequency $f_i$,  and discuss  detection of its GWs with  space interferometers.

From  the quadrupole formula, the chirp rate of its GW frequency is given by 
\begin{align}
 \frac{df}{dt} = \frac{96
 \pi^{8/3} G^{5/3} \mathcal{M}^{5/3} f^{11/3}}{5 c^5 } .
\end{align}
By integrating this equation, the remaining time before the merger is estimated to be 
\begin{align}
 t_\mathrm{mer}&= \frac{5 c^5 f_i^{-8/3}}{256 \pi^{8/3} G^{5/3}
                    \mathcal{M}^{5/3} }\label{tmer}\\
                     &=4{\rm yr} \left( \frac{f_i}{\SI{25}{mHz}} \right)^{-8/3}\left( \frac{\mathcal{M}}{15.5M_\odot}
 \right)^{-5/3}.
\end{align}
Similarly, after an observational period $T$ less than $t_{\rm mer}$, the final frequency $f_f$ is written by  
\begin{equation}
 f_f ( f_i , T ) = f_i\lmk \frac{t_{\rm mer}-T}{t_{\rm mer}}\rmk^{-3/8} .
\end{equation}
For $T>t_{\rm mer}$, we should formally put $f_f=\infty$ in our analysis below.  

The signal-to-noise ratio $\rho$ is given by
\begin{equation}
 \rho^2 = 4 \int_{f_i}^{f_f} \frac{| \tilde{h} (f) |^2}{(3/20)S(f)} df 
  \label{eq:snr}
\end{equation}
with the Fourier transformed amplitude $ \left| \tilde{h}(f) \right|$ and the angular averaged noise spectrum $S(f)$  for a triangular detector unit.  In Fig. 3, we show the target sensitivity $S(f)$ for LISA (Robson, Cornish, \& Liu 2019), Taiji (Wang et al. 2020) and TianQin (Huang et al. 2020), along with the Galactic confusion noise for $T=10$yr  (Robson, Cornish, \& Liu 2019).  For LISA with the armlength of 2.5Gm, the A- and E-channels  can be regarded as the main data streams for  BBH detection at $f\lsim $30mHz (Prince et al. 2002).

At the quadrupole order, the  GW amplitude is expressed as 
\begin{equation}
 \left| \tilde{h}(f) \right|^2 = \frac{5 G^{5/3} \mathcal{M}^{5/3}}{24
  \pi^{4/3} c^3 D^2 f^{7/3}} \times \frac{3}{4} \left[ F_+^2 \left(
                                                              \frac{1 +
                                                              \mu^2}{2}
                                                             \right)^2 +
  F_\times^2 \mu^2 \right] 
\end{equation}
with the distance $D$ and $\mu\equiv \cos I$ ($I$: the inclination angle).  The beam pattern functions $F_{+,\times}$ are defined for a triangular unit and normalized appropriately (in relation to the noise  spectrum $S(f)$). They are  generally time dependent. However, for LISA and Taiji, due to the annual rotation of their detector planes, we can replace them relatively accurately by their angular averages  
$\lla F_+^2\rra_a=\lla F_\times^2\rra_a=1/5$, at least for $t_{\rm mer}>T>2$yr (see e.g., Robson et al. 2019).

Then, for a given threshold $\rho_{\rm thr}$ of the signal-to-noise ratio, the effective volume of the BBH detection is estimated to be 
\begin{equation}
 V ( f_i , T ) = \frac{4 \pi}{3} \times U
  \frac{A^3}{\rho_\mathrm{thr}^3} I_7 ( f_i , T )^{3/2} .
  \label{eq:volume}
\end{equation}
Here, we put 
\begin{align}
 I_7 ( f_i , T ) & \equiv \int_{f_i}^{f_f} \frac{f^{-7/3}}{(3/20)S(f)}
 df  \label{int}\\
 A & \equiv \frac{G^{5/6} \mathcal{M}^{5/6}}{2^{3/2} \pi^{2/3} c^{3/2}}
\end{align}
and define 
\begin{equation}
 U\equiv \frac{1}{2} \int_{-1}^1 \left[ \left( \frac{1 + \mu^2}{2} \right)^2 +
                          \mu^2 \right]^{3/2} d \mu = 0.822 
 \label{eq:ave}
\end{equation}
for the inclination $\mu=\cos I$.  In this paper, unless otherwise stated, we put $\rho_\mathrm{thr}=10$ (see also \S 5).
In Fig. 3, around the optimal frequency band, the noise spectrum $S(f)$ of Taiji is $\sim2$ times smaller than that of LISA. Therefore,  the detectable volume  of Taiji is $2^{3/2}\sim3$ times larger.  

In fact, the detector plane of TianQin will be always normal to the direction of a white dwarf binary RX J0806.3+1527 (Luo et al. 2016). Then, the averaging for the beam pattern functions does not work as in the case of LISA and Taiji.  Using the prescription for the separation of the direction and orientation angles (Seto 2014), this difference increases  the detectable volume of TianQin {approximately} by a factor of 
\begin{eqnarray}
\frac{\lla ( F_+^2+F_\times^2)^{3/2} \rra_a}{\lla  F_+^2+F_\times^2 \rra_a^{3/2}}\approx \frac{U}{(4/5)^{3/2}}=1.15.\label{fct}
\end{eqnarray}
In order to include this correction, in Fig. 3,  we actually multiplied $1.15^{-2/3}\sim 0.91$ to the noise curve of TianQin. 

\section{expected detections}
In this section, we estimate the expected numbers of  detectable BBHs  with the proposed space interferometers.

\subsection{one component model}

We first discuss a simple model with the single chirp  mass $\mch={\bar \mch}_{10/3}=15.5\so$ and the comoving merger rate $R$ at $R_0= 19.1 {\rm Gpc^{-3} yr^{-1}}$.   
As explained later in the next subsection, by choosing this weighted average of the chrip masses, we can well  reproduce the low frequency profile obtained by a  more detailed analysis including  the chirp  mass spectrum. 

The expected detection number $dN/d\ln f_i$ per logarithmic  frequency interval is evaluated as  
\begin{equation}
 \frac{dN}{d \ln f_i} = V (f_i,T) \frac{dn}{d \ln f_i} , \label{eq:exact}
\end{equation}
in terms of the effective volume $ V (f_i,T) $ defined in Eq. (\ref{eq:volume}) and the frequency distribution of the BBHs $dn/d\ln f$.   Using the continuity equation in the frequency space, we have 
\begin{align}
 \frac{dn}{d \ln f} & = f\lmk  \frac{df}{dt}\rmk^{-1}  R = \frac{5 c^5 R}{96
 \pi^{8/3} G^{5/3} \mathcal{M}^{5/3} f^{8/3}} \label{eq:distrib} \\
 & = \SI{2.3e-6}{Mpc^{-3}} \notag \\
 & \times \left( \frac{f}{\SI{10}{\milli\hertz}} \right)^{-8/3} \left(
 \frac{\mathcal{M}}{15.5M_\odot} \right)^{-5/3} \left(
 \frac{R}{\SI{19.1}{Gpc^{-3}.yr^{-1}}} \right) .
\end{align}

In Fig. 4, we present our numerical results for the observational periods $T=4$ and 10yr with LISA and TianQin. The merging time becomes $t_{\rm mer}=4$ and 10yr respectively at the initial frequencies $f_i=25.0$ and  17.7mHz, as shown in Fig. 4.  At initial frequencies $f_i$ higher than these frequencies, the BBHs merge in observational periods $T$ (Sesana 2016). These binaries are important for ground-based detectors, and we specifically call them by merging BBHs below and let $N_{\rm mer}$ denote their expected   number.   Note that, for the single mass model,  we can make a clear identification of the merging BBHs in the frequency space, as in Fig. 4.   This is one of the reasons we use this simplified  model here. 

We define $N_{\rm tot}$ as the total number of expected BBH detections obtained by integrating Eq. (\ref{eq:exact}).
Because of the difference of the optimal bands, the number of merging BBHs $N_{\rm mer}$   is larger for TianQin than LISA.  However,  the total  numbers $N_{\rm tot}$ of the detectable BBHs are similar. Quantitatively,  we have  $N_{\rm tot}\sim 2$ and $\sim 9$ for $T=4$yr and 10yr, respectively.

For LISA and the observational period $T=10$yr, we also include the Galactic confusion noise spectrum and show the result with the thin solid curve in Fig. 4. We have a minor correction ($\sim 3\%$ for $N_{\rm tot}$) only  at the low frequency end.  Below, we ignore the Galactic confusion noise. 

At the low frequency regime, we have $t_{\rm mer}\gg T$ and the integral $I_7$ can be approximately evaluated as 
\begin{equation}
I_7(f_i,T)\simeq T  \lmk \frac{df}{dt}\rmk_{f_i} \frac{ f_i^{-7/3}}{(3/20) S(f_i)}.
\end{equation}
We then have 
\begin{equation}
 \frac{dN}{d \ln f_i} \approx \frac{\pi^{1/3}U G^{10/3}
  \mathcal{M}^{10/3} R}{3^{1/2} 5^{1/2} c^7 \rho_\mathrm{thr}^3}
  \frac{f_i^{-2/3} T^{3/2}}{[(3/20)S(f_i)]^{3/2}} \label{eq:mono},
\end{equation}
corresponding to the monochromatic frequency approximation in paper 1.  In Fig. 4, we present this expression with the dashed curves. As expected, they reproduce the thick solid curves well at $f_i\lsim 10$mHz.  

As shown in Fig. 4, most of the detectable BBHs have merger time shorter than $t_{\rm mer}\lsim 10^3$yr.
Then, for  $T\gsim 4$yr,   their chirp rates  ${\dot f_i}\sim f_i/t_{\rm mer}$ are much larger than its measurement error $\Delta {\dot f_i}\sim \rho^{-1}T^{-2}$  (see e.g., Takahashi \& Seto 2002; Toubiana et al. 2020).  Therefore, by using the observed chirp masses and distances,  we can safely distinguish extra-Galactic BBHs from numerous Galactic binaries.

\begin{figure}
 \includegraphics[width=.95\linewidth]{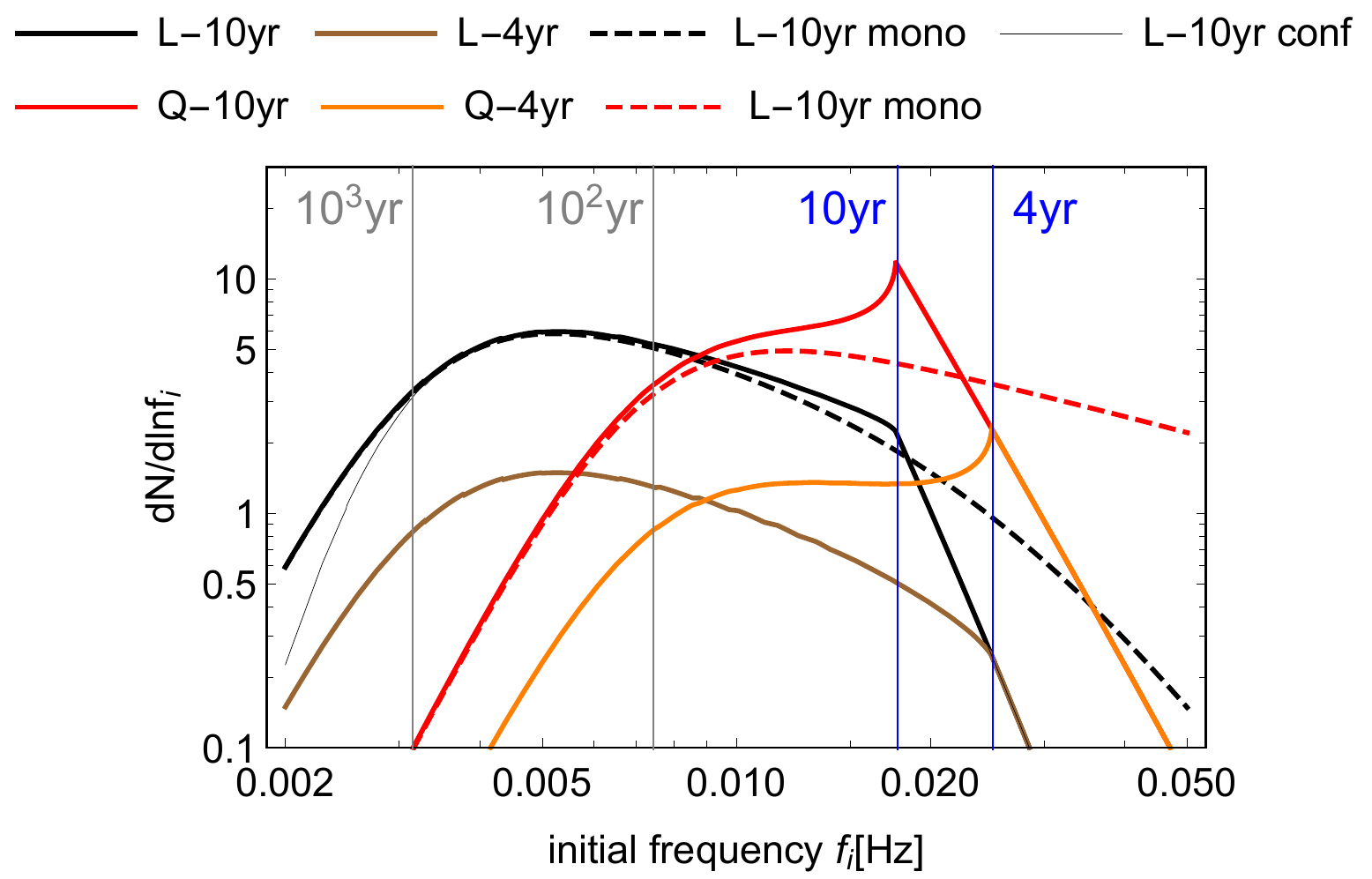} \caption{The expected number of  BBH detections  $dN/d\ln f_i$ per logarithmic initial frequency interval. We evaluated the simplified model with the single chirp mass $\mch=15.5\so$ and the comoving merger  rate $R=R_0=19.1 {\rm Gpc^{-3} yr^{-1}}$. The thick solid curves are for LISA and TianQin with observational periods $T=4$ and 10yr.  At frequencies  $f_i$ higher than  the blue vertical  lines, the BBHs merge in the observational periods $T$.  The dashed curves are the monochromatic approximation (\ref{eq:mono}) valid for $t_{\rm mer}\gg T$.   We include the Galactic confusion noise for the thin solid curve (LISA  10yr).   }  \label{fig:volume}
\end{figure}

\subsection{weighted results}

We now study the averaged results using Eq. (\ref{sl}) for the 62 BBHs. In Fig. 5, for $T=10$yr, we compare the results obtained with the full weighting  operation and the single mass model in the previous subsection.  We can see differences between them at $f_i\gsim 10$mHz (corresponding to $t_{\rm mer}=10$yr for  $\mch\sim  40\so$).

Quantitatively, in the case of  LISA, we have  $(N_{\rm tot},N_{\rm mer})=(8.9, 0.83)$  for the weighted calculation  and (9.1, 0.34) for the single mass model. The relative magnitude of $N_{\rm mer}$ might look inconsistent with the two black curves in  Fig. 5.  However, we should notice  that, for the weighted calculation,  the number  $N_{\rm mer}$  is dominated by heavier BBHs at the lower frequency range than the single mass model.

In Fig. 5, at $f_i\lsim7$mHz, the solid and dashed curves show reasonable agreement (less than 15\% differences). In this regime, the monochromatic approximation works well, and we have $dN/d\ln f_i\propto R \mch^{10/3}$ for the individual chirp masses as in Eq. (\ref{eq:mono}).  After the chirp mass weighting, we  have     $dN/d\ln f_i\propto R \lla  \mch^{10/3}\rra=R({\bar \mch}_{10/3})^{10/3}$.  Considering this relation, we have set $\mch={\bar \mch}_{10/3}=15.5\so$ for the single mass model in the previous subsection.

In Fig. 6, we compare the cumulative chirp mass distributions for various weights of BBHs.   Relative to the blue curve for the comoving merger rate (identical to the blue one in Fig. 2),  those detected by LVKC (listed in GWTC-3) have the weight $\mch^{5/2}$, as mentioned in \S 2.  {In fact, this should be regarded as the original observational data for our analysis. It has the identical vertical step of  1/62.} Under the monochromatic approximation (\ref{eq:mono}),  we have the stronger weight $\mch^{10/3}$ for LISA.   LISA is most likely  to detect BBHs with $\mch= 15-50\so$, somewhat larger than the GWTC-3  samples.    
We should notice that this mass range is higher than the weighted result ${\bar \mch}_{10/3}=15.5\so$, {partly due to the nonlinearity of the operation (\ref{mmnt}). }
%which basically works in the form of the  product $R ({\bar \mch}_{10/3})^{10/3}$ with the total
% comoving rate $R$.  
 At present, reflecting the limited number of the GWTC-3  samples, the higher mass end $\mch\gsim 50\so$  has a large statistical uncertainty, also indicated by the comparison with the solid curves. In the stellar mass  BBHs detected by LISA, a certain fraction $\sim 15\%$ could have chirp masses in the range $50-70\so$.

In Fig. 5, at $f_i\gsim 30$mHz, we can see a small deviation between the solid and dashed curves. For these BBHs, we have $T>t_{\rm mer}$ and the integral $I_7(f_i,T)$ becomes independent of the chirp mass [corresponding to $f_f=\infty$ in the integral (\ref{int})].  From Eq. (\ref{eq:exact}), we then have $dN/d\ln f_i\propto R  \mch^{5/6}$ for the individual masses and  $\propto  R\lla  \mch^{5/6}\rra=R({\bar \mch}_{5/6})^{5/6}$ after the weighting. Indeed, in Fig. 5,  at $f_i\gsim30$mHz, we find that the mismatch between the curves is  a factor of 
$\sim({\bar \mch_{5/6}}/{\bar \mch}_{10/3})^{5/6}=0.7$.

Next we discuss dependence of the expected numbers $(N_{\rm tot},N_{\rm mer})$ on the observational period $T$.  The nominal operation period of LISA is 4 yr (Amaro-Seoane et al. 2017).   In Fig. 7, we present our numerical results for LISA and TianQin.  For  $T=4$yr, we have $(N_{\rm tot},N_{\rm mer})=(2.3,\,0.10)$ with LISA and  (2.1, 0.61) with TianQin.  We will have a fair chance to detect extra-Galactic BBHs with LISA, but they are not likely to merge in the observational period.  TianQin has a larger chance to detect a merging BBH.   At $T=10$yr, we have  $(N_{\rm tot},N_{\rm mer})=(8.9, 0.83)$  and (7.8,\,3.1), respectively,  for LISA and TianQin.  {These numbers are basically consistent with the results of Liu et al. (2020).} 

As expected from Eq. (\ref{eq:mono}),  for LISA, the total number  $N_{\rm tot}$ is approximately proportional to $T^{3/2}$.  Meanwhile we can approximately show that $N_{\rm mer}\propto f_i(T)^{-14/3}\propto T^{7/4}$, ignoring the frequency dependence of the noise spectrum. As expected from Fig. 3, this approximation works better for TianQin. 

In any case,  the expected numbers become much  smaller than paper 1. For LISA, the detection number $N_{\rm tot}$ is dominated by nearly monochromatic BBHs and proportional to the product $R ({\bar \mch}_{10/3})^{10/3}$.  Comparing with paper 1, the product now becomes smaller by a factor of $(19.1/100)(15.5/28.6)^{10/3}=0.025$.  While the current target sensitivity of LISA is  better than some of these examined in paper 1, the renewal  of the astronomical information significantly reduced the expected numbers.

\begin{figure}
 \includegraphics[width=.95\linewidth]{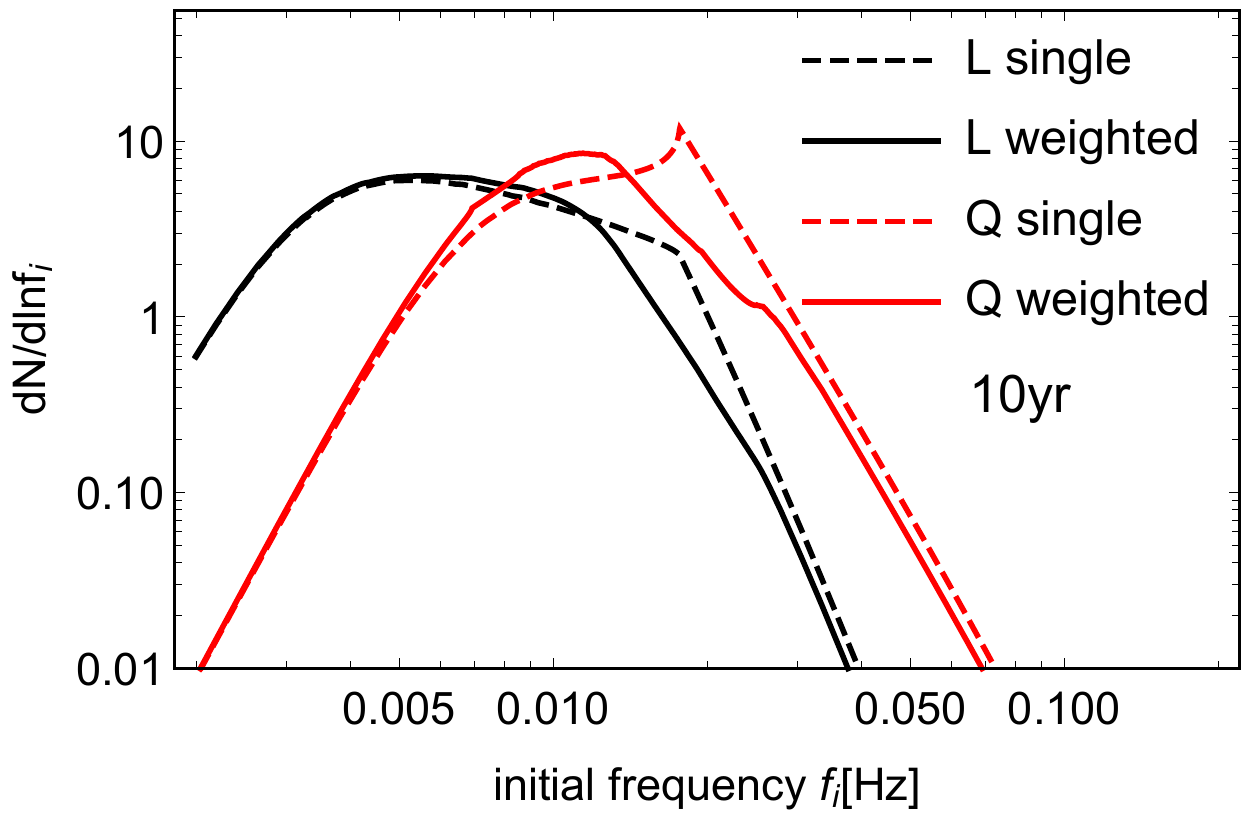} \caption{The comparison between the single mass calculation and the weighted results. At $f\lsim 7$mHz, their differences are less than $\sim 15\%$.   }  \label{fig:volume}
\end{figure}

\begin{figure}
 \includegraphics[width=.95\linewidth]{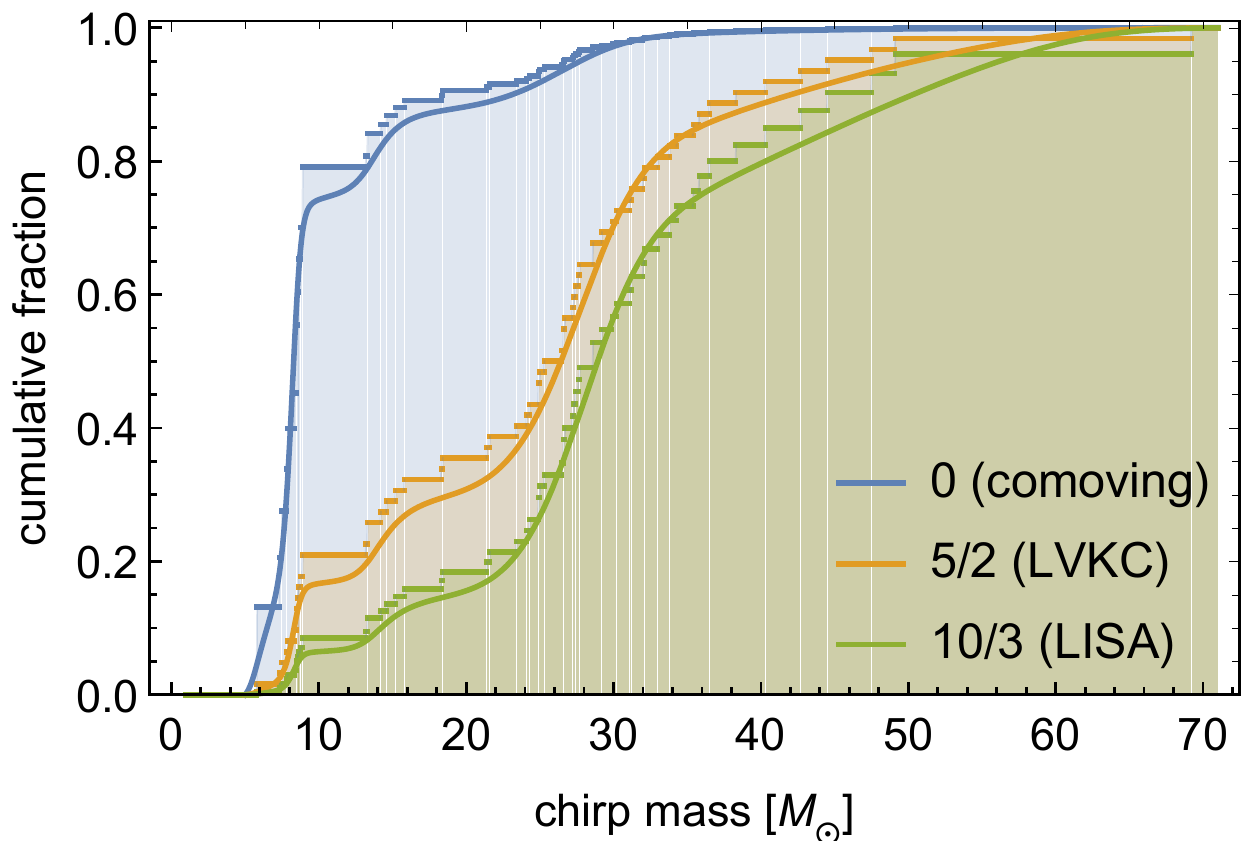} \caption{The cumulative chirp mass distributions for various weights of BBHs.  Relative to  the  comoving merger rate, the LVKC detection  (listed in GWTC-3) has the weight $\propto \mch^{5/2}$ and the LISA detection has  $\propto \mch^{10/3}$ under the monochromatic approximation. %We also plot the weight $\propto \mch^4$ for a reference. 
  {The solid curves are obtained with the FM model prediction in Abbott et al. (2021b). }}  \label{fig:volume}
\end{figure}

\begin{figure}
 \includegraphics[width=.95\linewidth]{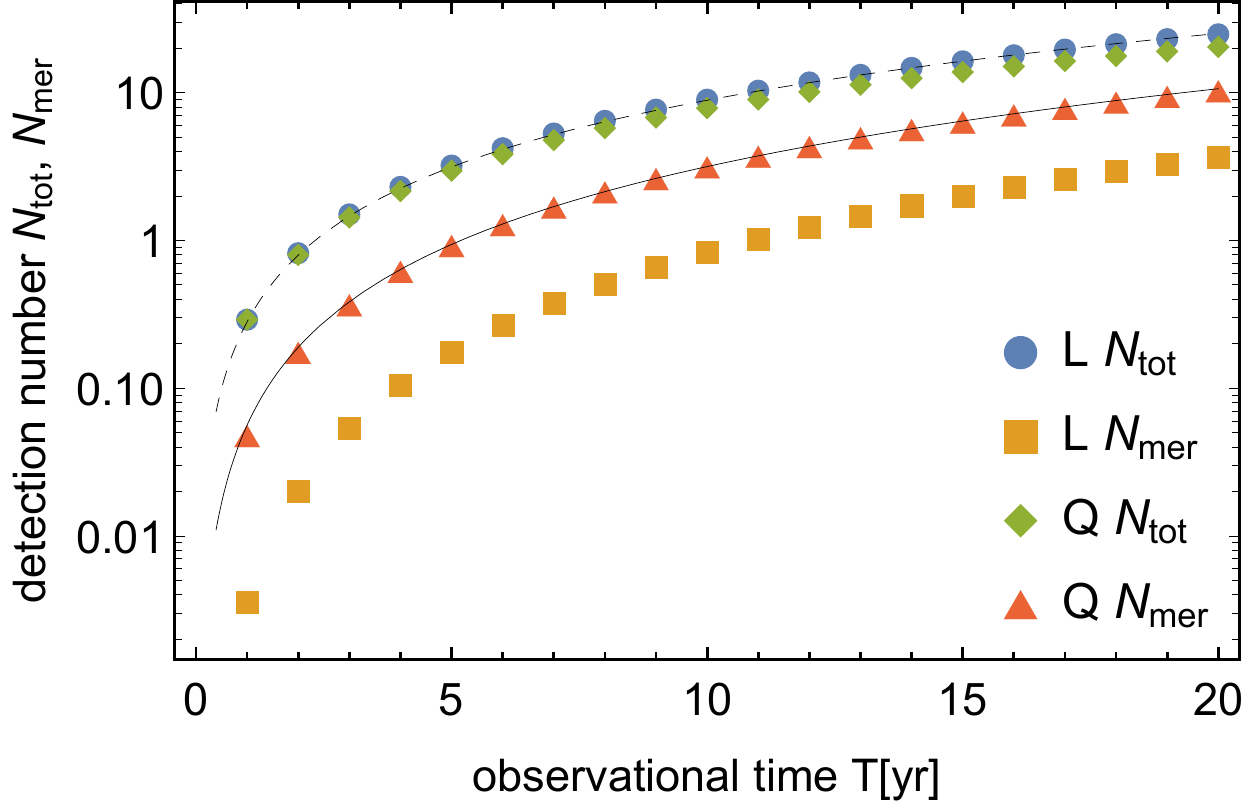} \caption{The expected detection numbers $N_{\rm tot}$ and $N_{\rm mer}$ for LISA and TianQin as functions of the observational periods $T$ (after weighting).   The dashed and solid curves are proportional to $T^{3/2}$ and $T^{7/4}$.}  \label{fig:volume}
\end{figure}

\subsection{joint operation}

Considering the timeline of the proposed missions such as LISA, Taiji and TianQin, we can expect their joint signal analysis.   
In this section, we discuss how the BBH detection could be improved by the coherent matched filtering analysis (see also Liu et al. 2020).  We specifically examine   the following  potential combinations of two 10yr-missions

(i) L2-10: two LISA-equivalent detectors operated   concurrently for 10 years (100\% time overlap),

(ii) L2-20:  two LISA-equivalent detectors operated  in a sequential order for a total of  20 years (0\% time overlap),

(iii) LQ-10:  LISA and TianQin operated concurrently for 10 years (100\% time overlap).

For L2-10 and L2-20, we imagine collaboration of  LISA and Taiji (ignoring the difference of their noise spectra). 
For evaluating the total signal-to-noise ratio of the coherent analysis, we simply added the frequency integrals $I_7$.  %keeping the correction factor for TianQin  mentioned around  Eq. (\ref{fct}).    
%We thus  ignored the subtle issue about the detectable volume for LQ-10

In Fig. 8, we present the frequency distributions  $dN/d\ln f_i$ for the three combinations, together with the previous results for LISA (L-10) and  TianQin (Q-10).  We can see  the asymptotic   convergences for the two pairs (LQ-10, L-10) and (L2-10, L2-20) at the low frequency regime, and similarly  (Q-10, LQ-10) at the high frequency regime.  We can easily understand these behaviors, {considering the optimal bands of detectors and the scaling relation of the monochromatic approximation (\ref{eq:mono}).  }  The asymptotic convergence is also observed for (L-10, L2-20) at the high frequency regime, because all the binaries merge within 10yr.

In Table 1, we show the total number $N_{\rm tot}$ at the end of the operations.   For the three networks L2-10, L2-20 and LQ-10, we have $N_{\rm tot}\sim 20$ that is $2^{3/2}\sim3$ times larger than L-10 and Q-10. 
In Table 1, we also present the number of merging BBHs  $N_{\rm mer}$ with the remaining times $t_{\rm mer}<10$yr and 20yr (slightly expanding the definition of $N_{\rm mer}$).  We should recall that the time $t_{\rm mer}$ is defined with the initial frequency $f_i$ at the the beginning of the observation [see Eq.  (\ref{tmer})].   If the results are compared at the same time span for  $t_{\rm mer}<$20yr, L2-10 has larger numbers of $(N_{\rm tot},N_{\rm mer})$ than L2-20. Thus, the coincident operation is more advantageous than the sequential operation for detecting stellar-mass BBHs.  {The coincident operation is also beneficial for  sky localization of merging massive black hole binaries (see e.g., Ruan et al. 2018; Wang et al. 2020) and for correlation analysis  of stochastic backgrounds (see e.g.,   Seto 2020).}

\if0
Table 1 also shows that Tainqin has an interesting tole to increase $N_{\rm mer}$, becase of its optimal band  higher than LISA and Taiji. 
\fi

\if0%%%%%%%%%%%%%%%%%%%%%%%%%%%%%%%
\begin{figure}
 \includegraphics[width=.95\linewidth]{ex.pdf} \caption{Prospects for  the potential combinations of space missions. The curves L-10yr and Q-10yr are the baselines results for separately operating LISA and TianQin for $10$yr.  LQ-10yr is  for concurrently using LISA and TianQin for 10yr.   L2-10yr and L-20yr  respectively show the results  for using  two LISA-equivalent missions   for  concurrently  10yr and  for totally  20 yr in a sequential order.  }  \label{fig:volume}
\end{figure}
\fi%%%%%%%%%%%%%%%%%%%%%%%%%%%%%%%%%%

\begin{figure}
 \includegraphics[width=.95\linewidth]{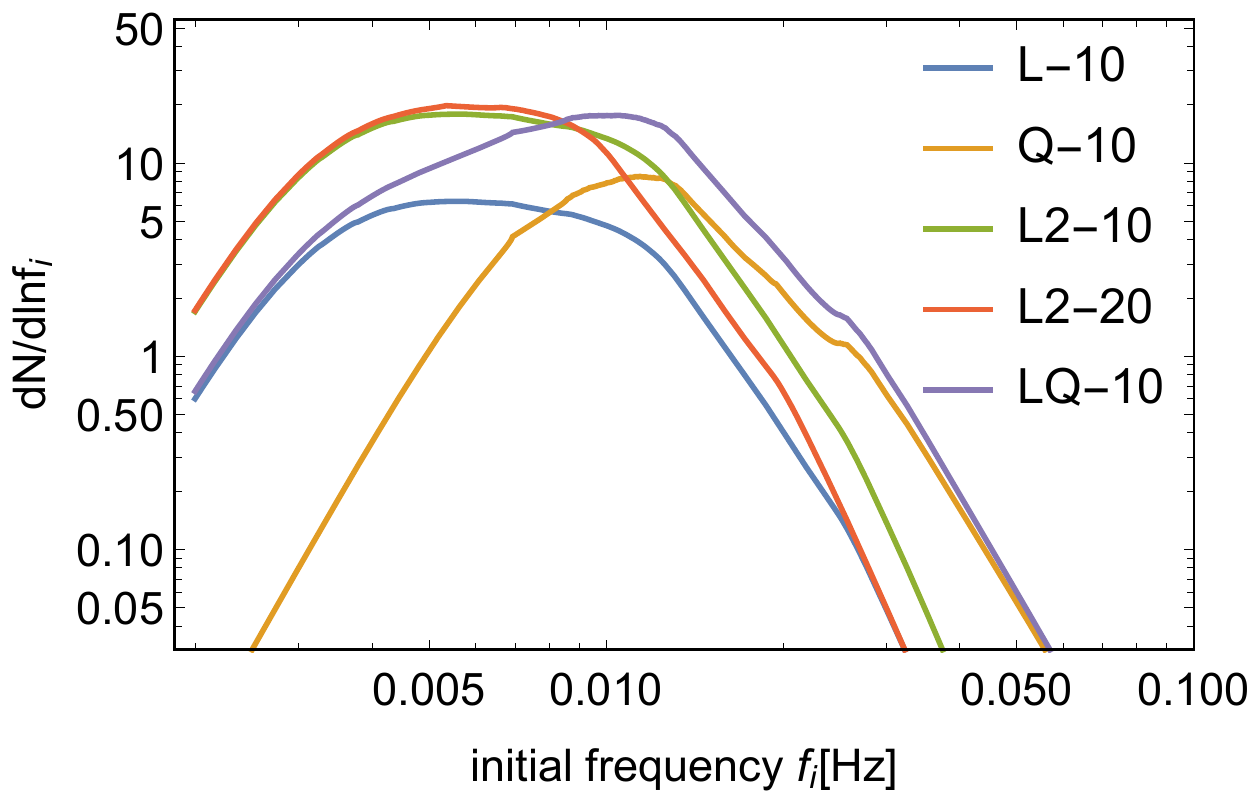} \caption{Results for the joint operations of space missions (see the main text for details of the five cases). }  \label{fig:volume}
\end{figure}

\begin{table}
\caption{The total detection numbers $N_{\rm tot}$ for the baseline operations (L-10 and Q-10), and the joint operations (L2-10, L2-20 and LQ-10).   We also show the number of detectable BBHs $N_{\rm mer}$ with the merger time $t_{\rm mer}$ less than 10 and 20yr. }
 \centering
\begin{tabular}{@{}l|lllll@{}}
\toprule
  &   L-10    & Q-10    & L2-10   & L2-20    & LQ-10    \\ \midrule
 $N_{\rm tot}$ & 8.9  & 7.8   &    25.3      &24.7  & 22.2     \\ \bottomrule
 
$N_{\rm mer}$ ($t_{\rm mer}<$10yr) & 0.83  & 3.1   &   2.3      & 0.83   &  5.2    \\

$N_{\rm mer}$ ($t_{\rm mer}<$20yr)  & $2.1 $ & $5.2$ & $5.9 $& 3.6 & $9.9 $ \\ \bottomrule
\end{tabular}
\end{table}

\if0 00000000000000000000000000000000000000000
% \usepackage{booktabs}
\begin{table}
\caption{The total detection numbers $N_{\rm tot}$ for the baseline operations (L-10 and Q-10), and the joint operations (L2-10, L2-20 and LQ-10).    }
 \centering
\begin{tabular}{@{}l|lllll@{}}
\toprule
  &   L-10    & Q-10    & L2-10   & L2-20    & LQ-10    \\ \midrule
 $N_{\rm tot}$ & 7.6   & 6.7   &    21.5      &21.1  & 18.9     \\ \bottomrule
\end{tabular}
\end{table}
\begin{table}
\caption{The number of detectable BBHs with the merger time $t_{\rm mer}$ less than 10 and 20yr.    }
 \centering
\begin{tabular}{@{}l|lllll@{}}
\toprule
     & L-10    & Q-10    & L2-10   & L2-20    & LQ-10    \\ \midrule
$N_{\rm mer}$ ($t_{\rm mer}<$10yr) & 0.70  & 2.7   &   2.0       & 0.70   &  4.4    \\

$N_{\rm mer}$ ($t_{\rm mer}<$20yr)  & $1.8 $ & $4.5$ & $5.0 $& 3.1 & $8.4 $ \\ \bottomrule
\end{tabular}
\end{table}
\fi%%%%%%%%%%%%%%%%%%%%%%%%%%%%%%%%%%%

\section{discussions}

So far, we have discussed BBH detection with space interferometers, setting the threshold at $\rho_{\rm thr}=10$.  In reality, for a fixed chirp mass, we generally  need a larger number of templates to find BBHs with space  interferometers than  ground based ones.  Therefore, the  required threshold   $\rho_{\rm thr}$ would be higher for the former and could be $\rho_{\rm thr}\sim 15$ (Moore et al.  2019).  If we simply put  $\rho_{\rm thr}=15$ (ignoring the frequency dependence) instead of 10, the detection numbers are reduced by a factor of $(15/10)^{-3}\sim0.3$ from our estimation. Meanwhile,  we should also notice that ground based interferometers will be able to detect  the same BBHs later at higher signal-to-noise ratios.  Thus, with multi-band observations for merging BBHs,  we can decrease the threshold  $\rho_{\rm thr}$ for  space interferometers (Wong et al. 2018).

For the spatial distribution of BBHs, we have assumed a smooth extra-Galactic component.  However, at the lower frequency regime (e.g., $f\sim1$mHz), the detectable distances become smaller, and we might need to evaluate BBHs in our local group (e.g., Seto 2016).  Unfortunately, at present, we observationally know little about  the merger rate and chirp mass spectrum in the local group (see also Lamberts et al. 2018; Wagg et al. 2021).

We have also assumed that BBHs have circular orbits.   Our results  will be modified only slightly for small residual eccentricities, e.g.,  $e  \sim 0.05$ at 5mHz (Nishizawa et al. 2017).  However,  the detectable numbers could be largely changed, if the BBHs typically  have higher eccentricities e.g., $e \sim 0.5$ in the  LISA band (see also Breivik et al. 2016; Samsing \& D'Orazio 2018;  Chen \& Amaro-Seoane 2017).

\section{summary}

In this paper, we updated the prospects of extra-Galactic BBH search with space interferometers such as LISA and TianQin.  We used the recent catalog GWTC-3 (Abbott et al. 2021b), targeting BBHs with component masses around 5-100$\so$ (and chirp masses $5-70\so$).  We directly incorporated the chirp masses distribution of the 62 BBHs detected at high significance.

For LISA, the total BBH detections are estimated to be $N_{\rm tot}\sim 2 (T/4{\rm yr})^{3/2}(\rho_{\rm thr}/10)^{-3}$ with the detection threshold $\rho_{\rm thr}$. 
LISA detection will be  dominated by nearly monochromatic BBHs in the mass range $\mch=$15-50$\so$ {with relatively large uncertainties above $50\so$. } 
 As for the dependence on astronomical information, we have the scaling relation $N_{\rm tot}\psim R({\bar \mch}_{10/3})^{10/3}$ with the comoving merger rate $R$ and an weighted average of the chirp masses ${\bar \mch}_{10/3}$. 
Compared with our previous estimation shortly after announcement of GW150914, the product  $N_{\rm tot}\psim R({\bar \mch}_{10/3})^{10/3}$  is reduced by a factor of  $\sim1/40$  [$\sim1/5$ from $R$ and $\sim 1/8$ from $({\bar \mch}_{10/3})^{10/3}$].

The Chinese project TianQin will have a total detection number $N_{\rm tot}$ similar to LISA.  Meanwhile,  it has potential to find $N_{\rm mer}\sim0.6  (T/4{\rm yr})^{7/4}(\rho_{\rm thr}/10)^{-3}$ BBHs that merge in the observational period $T$. Because of the difference of the optimal bands,   LISA will have   4-5 times smaller $N_{\rm mer}$.  Therefore, during its nominal operation period $T\sim4$yr, LISA alone is not likely to detect any  merging stellar-mass BBH, even optimistically counting an uncertainty factor of $\sim 2$ for the overall comoving merger rate.  A longer operation period and joint data analysis with other detectors can largely improve the prospects for the detection.

\section*{Acknowledgements}

This work is supported by JSPS Kakenhi Grant-in-Aid for  Scientific Research (Nos.~ 17H06358,
18H05236, 19K03870, 19K14720 and  20H00158).

\bibliographystyle{mn2e}

\end{document}